\documentclass[aps,prl,twocolumn,showpacs,floatfix]{revtex4}
\usepackage{amssymb}
\usepackage{amsmath}
\usepackage{graphicx}

\begin{document}

\title{Oscillations of magnetization in topological line-node semimetals}

\author{G.~P.~Mikitik}
\affiliation{B.~Verkin Institute for Low Temperature Physics \&
Engineering, Ukrainian Academy of Sciences, Kharkov 61103,
Ukraine}

\author{Yu.~V.~Sharlai}
\affiliation{B.~Verkin Institute for Low Temperature Physics \&
Engineering, Ukrainian Academy of Sciences, Kharkov 61103,
Ukraine}

\begin{abstract} We theoretically investigate the phase of the de Haas - van Alphen oscillations in topological line-node semimetals. In these semimetals the chemical potential of charge carriers can essentially depend on the magnetic field, and this dependence   changes the phase of the oscillations as compared to the phase in a    three-dimensional metal with a band-contact line. Our results elucidate recent experimental data on the Berry phase for certain   electron orbits in ZrSiS, ZrSiTe, and ZrSiSe.
\end{abstract}


\maketitle

\section{Introduction}

In recent years much attention has been given to the topological line-node semimetals in which the conduction and valence bands touch along lines in the Brillouin zone and disperse linearly in directions perpendicular to these lines \cite{hei1,balents,pie,weng,mullen,xie,kim,yu,yama,phil,schoop,
neupane1,schoop1,huang1,bian,chen}.
It is necessary to emphasize that the contact of the electron energy bands along the lines is the widespread phenomenon in crystals \cite{herring,m-sh14,kim,fang}. For example, such contacts of the bands occur even in many simple metals, and graphite \cite{graphite}, beryllium \cite{beryl}, aluminium \cite{al}, LaRhIn$_5$ \cite{prl04} are among them. However, the degeneracy energy of the bands, $\varepsilon_d$, is not constant along such lines, and this  energy $\varepsilon_d$ varies in the interval from its minimum $\varepsilon_{min}$ to its maximum $\varepsilon_{max}$ values. A   crystal with the band-contact line can be named the topological semimetal if the difference $\varepsilon_{max}- \varepsilon_{min}\equiv 2\Delta$ is sufficiently small
and if the chemical potential $\zeta$ of the electrons does not lie far away from the mean energy $\varepsilon_d^0 \equiv (\varepsilon_{max}+ \varepsilon_{min})/2$ of the line. Rhombohedral graphite \cite{pie,mcclure,kopnin}, Ca$_3$P$_2$ \cite{xie} and  CaAgP \cite{yama}, Cu$_3$NZn and Cu$_3$NPd \cite{kim,yu}, ZrSiS \cite{schoop,neupane1,chen}, ZrSiTe \cite{schoop1}, alkaline-earth germanides and silicides \cite{huang1}, PbTaSe$_2$ \cite{bian} are examples of the line-node semimetals.

The magnetization of electrons in a crystal with the band-contact line characterized by large $\Delta$ was theoretically investigated many years ago \cite{m-sv,m-sh}, and it was found
that the magnetic susceptibility of the electrons exhibits a giant anomaly when $\zeta$ approaches one of the energies  $\varepsilon_{min}$ or $\varepsilon_{max}$ which correspond to the points of the electron topological transitions of $3\frac{1}{2}$ kind \cite{m-sh14}. In the topological semimetals the interval $2\Delta$ is small, the critical energies $\varepsilon_{min}$ and $\varepsilon_{max}$ are close to each other, and the character of the anomaly in the susceptibility changes. The magnetic susceptibility in the case of the line-node semimetals was considered for weak magnetic fields in Ref. \cite{kosh15} and for arbitrary magnetic fields in Refs.~\cite{m-sh16,m-sh17}.

It is well known \cite{Abr,Sh} that at low temperatures the magnetization of electron in metals exhibits the de Haas - van Alphen oscillations. These oscillations are described by a periodic function of $cS_{ex}/(e\hbar H)-2\pi\gamma$ where $S_{ex}$ is the extremal cross-section area of the Fermi surface, and $\gamma$ is the constant in the semiclassical quantization rule. This $\gamma$ is expressed in terms of the Berry phase $\Phi_B$ for the electron orbit in the extremal cross section (see Eq.~(\ref{5}) below).
The characteristic feature of the topological line-node semimetals is that the de Haas - van Alphen oscillations are shifted in phase \cite{m-sh17} as compared to the case of metals for which the band-contact lines are absent, and $\Phi_B=0$, $\gamma=1/2$. The shift is due to the Berry phase $\pi$ for electron orbits surrounding the band-contact line \cite{prl}. Recently, the de Haas - van Alphen \cite{Hu1,Hu}, Shubnikov - de Haas \cite{Ali1,wang1,pez,singha}, and thermoelectric power \cite{matus} oscillations in magnetic fields were experimentally investigated in the line-node semimetals ZrSiS,  ZrSiTe, and  ZrSiSe, and intermediate values of the Berry phase (other than $0$ and $\pi$) were obtained for a number of the electron orbits. In this paper we suggest an explanation of these unusual values of $\Phi_B$ detected in the experiments.

Our explanation is based on the following considerations:
Due to small values of $\Delta$, the dispersion relation for the  electrons near the band-contact line is similar to the dispersion relation in layered metals \cite{m-sh17}. It is known \cite{champel,luk04,luk11,gus} that in such metals placed in the magnetic field $H$ a crossover from the three-dimensional electron spectrum to the quasi-two-dimensional one occurs with increasing $H$. In the case of the quasi-two-dimensional spectrum a dependence of the chemical potential on the magnetic field is strong \cite{Sh}, and this dependence changes the phase shift of the oscillations. We show that in the crossover region of the magnetic fields and in the region of the quasi-two-dimensional spectrum the shift can differ from $\pi$ and $0$, simulating the case of the  Berry phase deviating from these values.

\section{Formulas for magnetization}

To clarify the essence of the matter, we consider the simplest  band-contact line, assuming that it has the shape of a straight line in the quasi-momentum space ${\bf p}$, and that the electron  dispersion relation in the vicinity of the contact line of the two bands ``$c$'' and ``$v$'' has the form:
\begin{eqnarray}\label{1}
 \varepsilon_{c,v}\!\!&=&\!\varepsilon_d(p_3)\!\pm E_{c,v}, \\
 E_{c,v}^2\!\!&=&\!b_{11}p_1^2+b_{22}p_2^2, \nonumber
 \end{eqnarray}
where the $p_3$ axis coincides with the line; $\varepsilon_d(p_3)$ describes a dependence of the  degeneracy energy along the line (the $\varepsilon_{max}$ and $\varepsilon_{min}$ mentioned above are the maximum and minimum values of the function  $\varepsilon_d(p_3)$);  $b_{11}$ and $b_{22}$ are positive constants specifying the Dirac spectrum in the directions perpendicular to the line. Below we also use the simplest approximation for the periodic function $\varepsilon_d(p_3)$,
 \begin{eqnarray}\label{2}
  \varepsilon_d(p_3)=\Delta \cos\!\left(\!\frac{2\pi p_3}{L}\!\right)=\Delta\cos\!\left(\!\frac{p_3d}{\hbar}\!\right),
  \end{eqnarray}
where $L=2\pi\hbar/d$ is the length of the line in the Brillouin zone, and $d$ is the appropriate size of the unit cell of the crystal. Besides, we neglect the electron spin (but take into account the two-fold degeneracy of the electron energy bands in spin in the formulas  below), and consider the case of the zero temperature $T$ and of the magnetic field $H$ parallel to the line.

The Fermi surface corresponding to the dispersion relation (\ref{2}) is a corrugated cylinder when the chemical potential $\zeta$ lies outside the interval from $\varepsilon_{min}$ to $\varepsilon_{max}$. If $\zeta$ is inside the interval, the Fermi surface has a self-intersecting shape, and at $\zeta=\varepsilon_{min}$ or $\varepsilon_{max}$ the electron topological transitions of $3\frac{1}{2}$ kind  occur \cite{m-sh14}.

If the magnetic field $H$ is directed along the line, the electron  spectrum corresponding to the Hamiltonian (\ref{1}) has the form  \cite{m-sh}:
 \begin{eqnarray}\label{3}
\varepsilon_{c,v}^l(p_3)&=&\varepsilon_{d}(p_3) \pm \!\left(\frac{e\hbar\alpha H}{c}l\right)^{1/2}\!, \\
\alpha&=&\alpha(p_3)=2(b_{11}b_{22})^{1/2}, \nonumber
 \end{eqnarray}
where $l$ is a non-negative integer ($l=0$, $1$, \dots),
with the single Landau subband $l=0$ being shared between the branches ``$c$'' and ``$v$''. Interestingly, even for $l\sim 1$, the spectrum (\ref{3}) exactly coincides with that obtained from the semiclassical quantization rule,
 \begin{equation}\label{4}
 S(\varepsilon_l,p_3)=\frac{2\pi e \hbar H}{c}(l+\gamma ),
 \end{equation}
where $S(\varepsilon,p_3)= 2\pi[\varepsilon- \varepsilon_d(p_3)]^2/ \alpha$ is the area of the cross section of the isoenergetic surface by the plane perpendicular to the magnetic field and passing through the point with the coordinate $p_3$, the constant $\gamma$
is expressed in term of the Berry phase $\Phi_B$ for the appropriate electron orbit \cite{prl}:
 \begin{equation}\label{5}
 \gamma=\frac{1}{2}-\frac{\Phi_B}{2\pi},
 \end{equation}
and $\Phi_B=\pi$ in our case of the orbit surrounding the band-contact line.

The magnetization of electrons in a line-node semimetal was calculated in Ref.~\cite{m-sh17} at an arbitrary shape of its band-contact line. Using this result, we obtain the following expressions for the magnetization $M_3$ directed along the band-contact line being considered:
\begin{eqnarray}\label{6}
   M_3(\zeta,H)\!=\!\frac{1}{2\pi^2}\!\left(\frac{e}{\hbar c}\right)^{3/2}\!\!\!\alpha^{1/2}\!H^{1/2}\!\!
  \int_{0}^{L}\!\!\!\!\!dp_3 K(u),
  \end{eqnarray}
where the integration is carried out over this line; \begin{eqnarray}\label{7}
  K(u)\!=\! \frac{3}{2}\zeta(-\frac{1}{2},\![u]\!+\!1\!)+\sqrt{u}([u]+ \frac{1}{2}), \end{eqnarray}
$\zeta(s,a)$ is the Hurwitz zeta function,
 \begin{equation}\label{8}
 u=\frac{[\zeta-\varepsilon_d(p_3)]^2 c}{e\hbar  \alpha H}=\frac{cS(\zeta,p_3)}{2\pi e\hbar H}.
 \end{equation}
and $[u]$ is the integer part of $u$.

In the topological semimetals, charge carriers (electron and holes) are located near the band-contact line, and their chemical potential $\zeta$ generally depends on the magnetic field, $\zeta=\zeta(H)$. This dependence can be derived from the condition that the charge-carrier density $n(\zeta,H)$ does not vary with increasing $H$,
\begin{equation}\label{9}
n(\zeta,H)=n_0(\zeta_0),
\end{equation}
where $n_0$ and $\zeta_0$ are the density and the chemical potential at $H=0$. At $T=0$ the densities $n_0(\zeta_0)$ and $n(\zeta,H)$ are described by the following expressions \cite{m-sh17}:
\begin{eqnarray}\label{10}
 n_0(\zeta_0)\!\!\!&=&\!\!\!\frac{1}{2\pi^2\hbar^3\alpha} \!\!\int_{0}^{L}\!\!\!\!\!dp_3
(\zeta_0-\varepsilon_d(p_3))^2\sigma(\zeta_0- \varepsilon_d(p_3)),~~~ \\
 n(\zeta,H)\!\!\!&=&\!\!\!\frac{eH}{2\pi^2 \hbar^2c}\int_{0}^{L}\!\!\!\!\!dp_3
 (\frac{1}{2}+[u])\sigma(\zeta-\varepsilon_d(p_3)),
\label{11}
 \end{eqnarray}
where $\sigma(x)=1$ if $x>0$, and $-1$ if $x<0$. On calculating $\zeta(H)$ with Eqs.~(\ref{9})--(\ref{11}), one can find the magnetization as a function of $n_0$ or $\zeta_0$, inserting $\zeta(H)$ into Eqs.~(\ref{6}) -- (\ref{8}).

In analyzing the effect of $\zeta(H)$ on the phase of the de Haas - van Alphen oscillations, we shall plot the so-called Landau-level fan diagrams \cite{Sh} commonly used in treatments of the experimental data. At $T=0$ the periodic in $1/H$ magnetization of electrons in metals exhibits singularities (sharp maxima or minima \cite{shen}) when the lower or upper edge of the $l$th Landau subband crosses the Fermi level $\zeta$. In the semiclassical limit ($l\gg 1$) and {\it under the assumption that} $\zeta$ {\it is independent of} $H$, such crossings occur at the magnetic fields $H_l$ determined by Eq.~(\ref{4}):
 \begin{equation*}
 \frac{1}{H_l}=\frac{2\pi e \hbar}{cS_{ex}(\zeta)}(l+\gamma ),
 \end{equation*}
where $S_{ex}(\zeta)$ is the minimum or maximum value of  $S(\zeta,p_3)$ with respect to $p_3$. Thus, if the positions $H_l$ of the singularities are known, the constant $\gamma$ can be found with the Landau-level fan diagram: Plotting the Landau-level index $l$ versus $1/H_l$ and continuing the obtained straight line up to the intersection of this line with the $l$ axis in which $(1/H_l)=0$, the coordinate $-\tilde\gamma$ of the intersection enables one to obtain $\gamma$:  $\gamma= \tilde\gamma$. It is important to emphasize that if $\zeta$ lies in the energy region where the Dirac spectrum occurs, one can use $H_l$ with $l\sim 1$ in the construction of the fan diagrams since, as was mentioned above, the semiclassical spectrum resulting from formula (\ref{4}) coincides with the exact one given by Eq.~(\ref{3}) even at small $l$. Note also that this procedure of determining $\tilde\gamma$, which characterizes the phase of the de Haas - van Alphen oscillations, is applicable to the case when $\zeta$ depends on $H$, but as shall be demonstrated below, $\tilde\gamma$ thus extracted does not generally coincide with the constant $\gamma$ specifying the semiclassical quantization rule.

\section{Discussion}

The quantity $u$ defined by Eq.~(\ref{8}) changes along the nodal line between its minimal $u_{min}$ and maximal $u_{max}$ values which  correspond to the minimal $S_{min}$ and maximal $S_{max}$ values of $S(\zeta,p_3)$ with respect to $p_3$. In the case of sufficiently weak magnetic fields when $u_{min}$, $u_{max}$, $u_{max}-u_{min}\gg 1$, i.e., when
\begin{eqnarray} \label{12}
 \frac{2\pi e\hbar H}{c}\ll S_{max},S_{min},S_{max}-S_{min},
  \end{eqnarray}
the Landau subbands $\varepsilon^l_{c,v}(p_3)$ with different $l$  overlap as in three-dimensional metals. According to Ref.~\cite{m-sh17}, in this case
formula (\ref{6}) reduces  to the well-known expression \cite{Sh,luk04,gus} describing the de Haas - van Alphen oscillations in a three-dimensional metal but with $\gamma=0$. For $\varepsilon_d(p_3)$ given by Eq.~(\ref{2}), this expression is a superposition of two periodic functions  determined by the two extremal cross-section areas $S_{min}=S(\zeta,p_3=0)$ and
$S_{max}=S(\zeta,p_3=L/2)$. The dependence $\zeta(H)$ is sufficiently weak in this three-dimensional case and practically has no effect on the oscillations \cite{Sh}.

Consider now stronger magnetic fields than in the case of    inequalities (\ref{12}). If $|\zeta|\gg \Delta$, the difference $u_{max}-u_{min}$ becomes less than unity when $u_{min}$ and $u_{max}$ are still large. In this situation,
the spectrum (\ref{2}) transforms into the spectrum of a quasi-two-dimensional electron system since the different Landau subbands $\varepsilon_{c,v}^l(p_3)$ do not overlap, and they look like broadened Landau levels for which the spacing between the nearest  Landau subbands in the vicinity of the Fermi level is larger than their width $2\Delta$. At $u_{max}-u_{min}\ll 1$, the quantity $u$  is practically  independent of $p_3$ running the line, $S(\zeta,p_3)\approx S(\zeta)$, the corrugation of the Fermi surface becomes unimportant, and formula (\ref{6}) is simplified as follows:
\begin{eqnarray}\label{13}
   M_3(\zeta,H)\!\approx\!\frac{1}{\pi d\hbar^{1/2}}\!\left(\frac{e}{c}\right)^{3/2}
   \!\!\!\alpha^{1/2}\!H^{1/2} K(u),
  \end{eqnarray}
where
\begin{eqnarray}\label{14}
   u\approx \frac{\zeta^2 c}{e\hbar  \alpha H}=\frac{cS(\zeta)}{2\pi e\hbar H},
  \end{eqnarray}
and the function $K(u)$ at large $u$ has the form \cite{m-sh17}:
\begin{equation}\label{15}
 K(u)\approx \frac{\sqrt{u}}{2}(u-[u]-0.5)=-\frac{\sqrt{u}}{2\pi}\sum_{l=1}^{\infty} \frac{\sin(2\pi lu)}{l}.
 \end{equation}
Equations (\ref{13})-(\ref{15}) describe saw-tooth oscillations of $M_3$ with changing  $1/H$, and they coincide with the appropriate expression \cite{luk04,luk11,gus} for a two-dimensional metal with the Dirac spectrum. A refined analysis of Eq.~(\ref{6}) at $|\zeta|\gg \Delta$ and $u_{max}$, $u_{min}\gg 1$ gives
\begin{equation}\label{16}
M_3\!\approx\!-\frac{e|\zeta|}{2 \pi^2 \hbar cd}\!\sum_{l=1}^{\infty}\! \frac{1}{l} \sin (2\pi l \frac{c(\zeta^2\!\!+\!\Delta^2)}{e \hbar \alpha H})J_0(4\pi l \frac{c \zeta\Delta }{e \hbar \alpha H})\,,~~
\end{equation}
where $J_0(x)$ is the Bessel function for which one has $J_0(x)\approx 1$ at $x\ll 1$ and $J_0(x)\approx (2/\pi x)^{1/2}\cos(x-\pi/4)$ at $x\gg 1$. Formula (\ref{16}) agrees with the appropriate expression of Refs.~\cite{luk04,gus} and reproduces both Eqs.~(\ref{13}) - (\ref{15}) at $u_{max}-u_{min}=4c \zeta\Delta /e \hbar \alpha H\ll 1$ and the formula for the de Haas-van Alphen oscillations in three-dimensional metals with $\gamma=0$ at $u_{max}-u_{min}\gg 1$.

\begin{figure}[tbp] 
 \centering  \vspace{+9 pt}
\includegraphics[scale=.90]{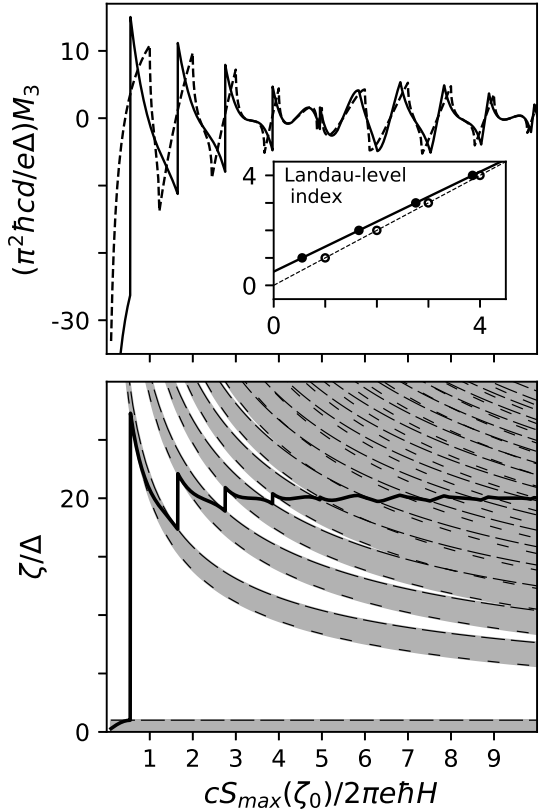}
\caption{\label{fig1} Top: The magnetization $M_3$, Eq.~(\ref{6}),  versus $1/H$ at fixed chemical potential $\zeta=\zeta_0=20\Delta$ (the dashed line) and at $\zeta(H)$ shown in the lower panel (the solid line). Here  $S_{max}(\zeta_0)=2\pi(\zeta_0+\Delta)^2/\alpha$, and $cS_{max}(\zeta_0)/(2\pi e\hbar H_{cr})\sim 5$. The inset depicts  the Landau-level indexes $l$ versus $cS_{max}(\zeta_0)/(2\pi e\hbar H_l)$; $1/H_l$ are the positions of the maxima of the functions $M_3(1/H)$ shown by the solid and dashed curves in the main plot. Bottom: The dependence of $\zeta$ on $1/H$ calculated with Eqs.~(\ref{9})--(\ref{11}) at $\zeta_0=20\Delta$. We also mark the Landau subbands by the dark background, and the short and long dashes indicate the lower and the upper edges of these subbands, respectively.
 } \end{figure}   

The crossover from the three-dimensional electron spectrum to the quasi-two-dimensional one occurs at the magnetic field $H_{cr}$ defined by the condition $u_{max}-u_{min}\sim 1$,
\begin{eqnarray}\label{17}
   H_{cr}\sim \frac{c(S_{max}-S_{min})}{2\pi e\hbar}=\frac{4c|\zeta|\Delta }{e\hbar  \alpha }.
  \end{eqnarray}
For $H\sim H_{cr}$, the spacing $\Delta\varepsilon_H$ between the Landau subbands in the vicinity of the Fermi level becomes comparable with their width $2\Delta$. Thus, the quasi-two-dimensional regime of the oscillations takes place in the interval of the magnetic fields $H_{cr}< H <H_1$ where $H_1$ is the field of the ultra-quantum limit,
\begin{eqnarray}\label{18}
   H_{1}\sim \frac{cS(\zeta) }{2\pi e\hbar  }=\frac{c\zeta^2 }{e\hbar  \alpha },
  \end{eqnarray}
at which $\Delta\varepsilon_H$ reaches $\zeta$, and the oscillations of the magnetization disappears. When $H$ changes in this interval, the chemical potential $\zeta(H)$ moves together with one of the Landau subbands, and then, at a certain value of $H$, it jumps from this subband to the neighboring one \cite{Sh}, Fig.~\ref{fig1}. This strong dependence $\zeta(H)$ noticeably changes the shape of the de Haas - van Alphen oscillations and can mask the correct values of $\gamma$ (and of the Berry phase) when $\gamma$ is found with the Landau-level fan diagram. Indeed, the jumps can occur at the fields $H_l$ for which $n(\zeta)$ in Eq.~(\ref{11}) becomes independent of $\zeta$. This situation is realized when $[u]$ in the right hand side of Eq.~(\ref{11}) is one and the same integer along the whole line. Let us denote this integer as $(l-1)$. Then, Eq.~(\ref{9}) takes the form:
\begin{equation}\label{19}
\frac{1}{H_l}=\frac{eL}{2\pi^2\hbar^2cn_0(\zeta_0)}\left (l-\frac{1}{2}\right ).
\end{equation}
These $H_l$ also mark the singularities in the magnetization since
immediately above and below $H_l$ the edges of the Landau subband touch the chemical potential, Fig.~\ref{fig1}. It follows from equation (\ref{19}) that the dependence of $1/H_l$ on $l$ is a straight line that intersects the $l$ axis at $l=-\tilde\gamma=1/2$, i.e, the Landau-level fan diagram plotted with the fields $H_l$
looks like in the case when $\gamma=1/2$ and the Berry phase $\Phi_B$ is equal to zero. (The value of $\tilde\gamma$ is defined up to an integer, and so $\tilde\gamma=-1/2$ and $\tilde\gamma=1/2$ are equivalent.) However, in reality one has $\Phi_B=\pi$, $\gamma=0$,  and the phase shift $\tilde\gamma$ extracted from the oscillations in the quasi-two-dimensional regime does not permit one to find $\Phi_B$ since $\gamma\neq \tilde\gamma$ now. The foregoing considerations are illustrated in Fig.~\ref{fig1} for which $\zeta_0/\Delta=20$ and $cS_{max}(\zeta_0)/(2\pi e\hbar H_{cr})\sim H_1/H_{cr}\sim 5$.

\begin{figure}[tbp] 
 \centering  \vspace{+9 pt}
\includegraphics[scale=.90]{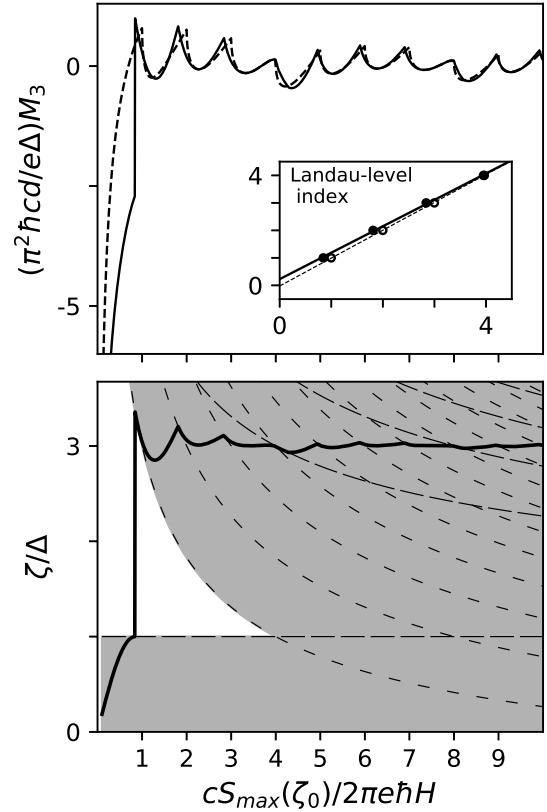}
\caption{\label{fig2} Top: The magnetization $M_3$, Eq.~(\ref{6}),  versus $1/H$ at fixed chemical potential $\zeta=\zeta_0=3\Delta$ (the dashed line) and at $\zeta(H)$ shown in the lower panel (the solid line). Here  $cS_{max}(\zeta_0)/(2\pi e\hbar H_{cr})\sim 1$. Bottom: The dependence of $\zeta$ on $1/H$ calculated with Eqs.~(\ref{9})--(\ref{11}) at $\zeta_0=3\Delta$. All the notations and the inset are similar to Fig.~\ref{fig1}. The intercept of the solid straight line in the inset gives $-\tilde\gamma=0.23\pm 0.04$. } \end{figure}   

Consider now the case when $\zeta$ is of the order of $\Delta$. In this situation $H_1 \sim H_{cr}$, and the crossover occurs near the ultra-quantum limit. Since, as was explained above, we have  $\tilde\gamma=0$ and $-1/2$ in the three-dimensional and in the quasi-two-dimensional regimes of the oscillations, respectively, one may expect that $\tilde\gamma$ takes intermediate values if this quantity is found in the crossover region. In Fig.~\ref{fig2} we show the de Haas - van Alphen oscillations of $M_3$ calculated with Eq.~(\ref{6}) at $\zeta_0/\Delta=3$ when $cS_{max}(\zeta_0)/(2\pi e\hbar H_{cr})\sim H_1/H_{cr}\sim 1$. In this situation we find $\tilde\gamma=-0.23\pm 0.04$ if the $H$-dependence of $\zeta$ is taken into account. Note that a relatively small contribution of the minimal cross section $S_{min}=S_{max}/4$ into the oscillations is also visible in the figure. This minimal cross section slightly affects the maxima in $M_3$ associated with $S_{max}$, and so we find $\tilde\gamma=0.02\pm 0.01$ even for the oscillations calculated at fixed $\zeta=\zeta_0$.

\begin{figure}[tbp] 
 \centering  \vspace{+9 pt}
\includegraphics[scale=.90]{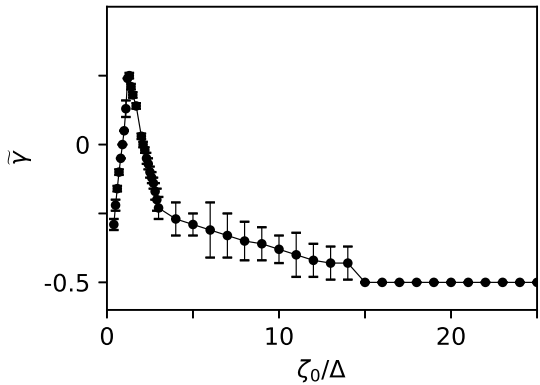}
\caption{\label{fig3} Dependence of $\tilde\gamma$ on $\zeta_0$. The values of $\tilde\gamma$ have been found with the Landau-level fan diagrams plotted using the first four sharp peaks in the functions $M_3(1/H)$.}
\end{figure}   

In Fig.~\ref{fig3} we present the dependence of $\tilde\gamma$ on $\zeta_0$. In the construction of this figure, values of $\tilde\gamma$ have been obtained, using the first four sharp maxima in the calculated functions $M_3(1/H)$. The small jump in $\tilde\gamma$ at  $\zeta_0/\Delta\approx 15$ is due to that at $\zeta_0/\Delta>15$  all the four maxima in $M_3$ are determined by the jumps in $\zeta$, whereas at $\zeta_0/\Delta<15$ the chemical potential is continues for a part of these maxima, cf. Figs.~\ref{fig1} and \ref{fig2}. At $\zeta_0/\Delta\approx 1$ the jumps in $\zeta$ disappear completely. For $\zeta_0/\Delta<3$ one has $S_{max}/S_{min}>4$, i.e., the ratio of the oscillation periods corresponding to the minimal and maximal cross sections is larger than $4$, and the effect of the low-frequency oscillations on the first four peaks in $M_3$ decreases. For this reason the break appears in the dependence $\tilde\gamma(\zeta_0)$ at $\zeta_0/\Delta\approx 3$. In other words, this break as well as the jump at $\zeta_0/\Delta \approx 15$ are caused by the relatively small number of the peaks in $M_3$ used in our plotting the Landau-level fan diagrams.

\begin{figure}[tbp] 
 \centering  \vspace{+9 pt}
\includegraphics[scale=.90]{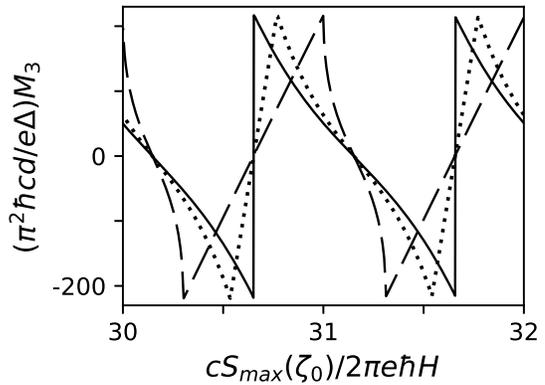}
\caption{\label{fig4} The magnetization $M_3$, Eq.~(\ref{6}),  versus $1/H$ at fixed chemical potential $\zeta=\zeta_0=400\Delta$ (the dashed line), at $\zeta(H)$ calculated with Eqs.~(\ref{9})--(\ref{11}) for $\zeta_0=400\Delta$ (the solid line), and at $\zeta(H)$ calculated with Eqs.~(\ref{10}), (\ref{11}), (\ref{20}), and (\ref{21}) for $\zeta_0=400\Delta$ and $\lambda=1$ (the dotted line). Here  $cS_{max}(\zeta_0)/(2\pi e\hbar H_{cr})\sim 100$.  All the notations are similar to Fig.~\ref{fig1}. The Landau-level fan diagram plotted using the sharp maxima of the dotted line in the interval between $cS_{max}(\zeta_0)/(2\pi e\hbar H)=20$ and $40$ gives  $\tilde\gamma=-0.33\pm 0.04$.}
\end{figure}

Consider now the situation when apart from the charge carriers located near the nodal line under study, there is an additional electron group in the semimetal. In particular, this situation occurs in ZrSiS. For simplicity, we shall neglect the quantization of electron energy in the magnetic fields for the charge carriers of this additional group. Note that this simplifying assumption is easily realized  even at low temperatures if the  cyclotron mass of these carries is essentially larger than the cyclotron mass of electrons near the nodal line. Then, with the additional electron group, equation (\ref{9}) is modified as follows:
\begin{equation}\label{20}
n(\zeta,H)+\frac{dN(\zeta_0)}{d\zeta_0}(\zeta-\zeta_0)=n_0(\zeta_0),
\end{equation}
where $dN(\zeta_0)/d\zeta_0$ is the density of the electron states of the additional group at the Fermi level. We shall specify this  density of the states by the formula,
\begin{equation}\label{21}
\frac{dN(\zeta_0)}{d\zeta_0}=\lambda\frac{n_0(\zeta_0)}{\zeta_0},
\end{equation}
where $\lambda$ is the dimensionless parameter. If $\lambda\to 0$, we return to the case of the single electron group located near the nodal line. As was shown above, in this case one always has $\tilde\gamma=-1/2$ for the oscillations in the  quasi-two-dimensional regime. If $\lambda\gg 1$, we arrive at the case of the constant chemical potential which is stabilized by the large additional electron group. In this case $\tilde\gamma=\gamma=0$ in the  quasi-two-dimensional regime. Hence, it is reasonable to expect that $\tilde\gamma$ will take intermediate values if $\lambda\sim 1$. In Fig.~\ref{fig4}, we show the oscillation of $M_3$ in the quasi-two-dimensional regime for three values of the parameter $\lambda=0$, $1$, $\infty$. It is seen that the phase of the oscillations at $\lambda=1$ indeed has a value which lies between the values corresponding to the other two cases. Thus, the intermediate values of $\tilde\gamma$ can be found not only in the crossover region but also in the quasi-two-dimensional regime of the oscillations if there is an addition group of charge carriers in the semimetal.

\section{Conclusions}

We theoretically investigate the phase of the de Haas -van Alphen oscillations in topological line-node semimetals, using the simple model for their nodal lines, Eqs.~(\ref{1}) and (\ref{2}). There are   two regimes of the oscillations. These regimes are determined by the relation between spacing $\Delta\varepsilon_H$ separating the Landau subbands in the vicinity of the Fermi level and the width of these subbands, $2\Delta$, resulting from the dispersion of the degeneracy energy along the nodal line.

For not-too-strong magnetic fields when $\Delta\varepsilon_H \ll 2\Delta$, the three-dimensional regime of the oscillations occurs, the dependence of the chemical potential $\zeta$ on the magnetic field $H$ is weak, and the constant $\tilde\gamma$ defining the phase of the oscillations coincides with the constant $\gamma$ in the semiclassical quantization rule, i.e., $\tilde\gamma=\gamma$.
Since the Berry phase $\Phi_B=\pi$ and, according to Eq.~(\ref{5}), $\gamma=0$ for the electron orbits surrounding the nodal lines, one can detect these lines, measuring the phase of the de Haas - van Alphen  oscillations in this case.

With increasing magnetic fields, the spacing between the Landau subband becomes larger than $2\Delta$, and the quasi-two-dimensional regime of the oscillations takes place. In this regime the dependence of $\zeta$ on $H$ is strong. This dependence changes the shape and the phase of the oscillations, and $\tilde\gamma \neq \gamma$ in this regime. At low temperatures ($T\ll \Delta\varepsilon_H$) we find that $|\tilde\gamma|=1/2$ for the extremal cross sections for which $\gamma=0$. Thus, the results of the phase measurements will imitate the case $\gamma=1/2$ and will not permit one to find the true values of $\gamma$ and of the Berry phase.

Due to the experimental data of Refs.~\cite{Hu1,Hu,Ali1,wang1,pez,singha,matus}, the special attention in our paper is given to the  situations in which values of $|\tilde\gamma|$ can be intermediate between $0$ and $1/2$. We show that these situations can occur in the region of the magnetic fields where the crossover from the three-dimensional regime to the quasi-two-dimensional one takes place, and in the quasi-two-dimensional regime if there is an additional group of charge carriers in the semimetal. In these cases, measurements of $\tilde\gamma$ (the phase of the oscillations) provide information on the electron energy spectrum of the semimetal, see Fig.~\ref{fig3}, rather than on the Berry phase of the electron orbits.

\end{document}